# Doping dependence of the specific heat of single crystal BaFe$_2$(As$_{1-x}$P$_x$)$_2$

C. Chaparro[1], L. Fang[1], H. Claus[1], A. Rydh[2], G. W. Crabtree[1,3], V. Stanev[1], W. K. Kwok[1], U. Welp[1]

[1] Materials Science Division, Argonne National Laboratory, Argonne, Illinois 60439, USA

[2] Department of Physics, Stockholm University, SE-10691 Stockholm, Sweden

[3] Department of Physics, University of Illinois at Chicago, Chicago, Illinois 60607, USA

We present specific heat measurements on a series of BaFe$_2$(As$_{1-x}$P$_x$)$_2$ single crystals with phosphorous doping ranging from x = 0.3 to 0.55. Our results reveal that BaFe$_2$(As$_{1-x}$P$_x$)$_2$ follows the scaling $\Delta C/T_c \sim T_c^2$ remarkably well. The clean-limit nature of this material imposes new restraints on theories aimed at explaining the scaling. Furthermore, we find that the Ginzburg-Landau parameter decreases significantly with doping whereas the superconducting anisotropy is $\Gamma \sim 2.6$, independent of doping.

Empirical relations expressing universal trends in the behavior of classes of materials have proven important in clarifying underlying physical mechanisms. Examples include, among others, the Kadowaki-Woods relation [1] relating the electron effective mass enhancement to the temperature dependence of the resistivity in Fermi liquid systems, and the Uemura plot [2] and Homes scaling [3] establishing relations between the superfluid density and the value of $T_c$ in superconductors. Recently, Bud'ko, Ni and Canfield [4] observed that the specific heat anomaly at the superconducting transition of a series of Ba(Fe$_{1-x}$Co$_x$)$_2$As$_2$ and Ba(Fe$_{1-x}$Ni$_x$)$_2$As$_2$ crystals displayed an unexpected scaling of the form $\Delta C/T_c \sim T_c^2$ (BNC-scaling). Since, this scaling has been found to apply in a wide variety of Fe-pnictide and Fe-chalcogenide superconductors, see compilations in [5, 6] and also Fig. 5 below. However, a completely satisfying theoretical explanation is still lacking.

Here, we present specific heat measurements on a series of BaFe$_2$(As$_{1-x}$P$_x$)$_2$ single crystals with phosphorous content ranging from near optimum doped x = 0.3 to strongly over-doped x = 0.55. In contrast to most other members of the Fe-pnictide and Fe-chalcogenide superconductors, BaFe$_2$(As$_{1-x}$P$_x$)$_2$ can be grown with very high purity as evidenced by the observation of de Haas van Alphen measurements giving detailed knowledge on the electronic structure [7, 8], and by the low value of critical current density [9]. Furthermore, thermal conductivity [10], NMR [11], penetration depth [12] and ARPES [13] measurements indicate the presence of line nodes in the superconducting gap although the detailed gap structure is still controversial. Our specific heat results reveal that BaFe$_2$(As$_{1-x}$P$_x$)$_2$ follows the BNC-scaling remarkably well. The high purity of this material imposes new restraints on theories aimed at explaining the scaling since models based on specific electron scattering mechanisms appear to be inconsistent with our results. We find that the Ginzburg-Landau parameter $\kappa_c$ decreases significantly with doping, whereas the superconducting anisotropy is $\Gamma$~2.6, independent of doping, indicative of the dominating role of the electron Fermi surface sheets in forming the superconducting state.

High purity $BaFe_2(As_{1-x}P_x)_2$ single crystals were grown using a self flux method. Thoroughly mixed high purity Ba flakes (99.99%, Aldrich) and FeAs and FeP (homemade from Fe, As and P, 99.99%, Aldrich) powders were placed in $Al_2O_3$ crucibles. The crucible was sealed in an evacuated quartz tube under vacuum and heated up to 1180 $^o$C and cooled down to 900 $^o$C with a rate 2 $^o$C/min. This procedure yielded hundreds of crystals with platelet shape. The largest size is ~1 mm, with most crystals ranging in size between 300~600 µm. The concentration of phosphorus was determined using X-ray energy dispersive spectra (EDS). Low-field magnetization measurements in a field of 1 Gauss were performed in a custom-built SQUID magnetometer [14]. The measurements were performed on warming after initially cooling in zero field. We performed the caloric measurements using a membrane-based steady-state *ac*-micro-calorimeter [15]. It utilizes a thermocouple thermometer composed of Au-1.7% Co and Cu films deposited onto a 150 nm thick $Si_3N_4$ membrane. The current calorimeter works for temperatures above ~6K. The crystals were mounted onto the thermocouple with minute amounts of Apiezon N grease. An *ac*-heater current at a frequency of typically 47 Hz is adjusted such as to induce 50-200 mK oscillations of the sample temperature.

The lower inset of Fig. 1 shows the temperature dependence of the magnetization of the $BaFe_2(As_{1-x}P_x)_2$ single crystals used in this study. The magnetic transition widths are generally less than 1 K, with the sharpest transition being only 200 mK wide, underlining the high quality of the crystals. The top inset of Fig. 1 shows a picture of a cleaved single crystal displaying a flat and shiny surface. The main panel shows the heat capacity of the x = 0.3 - crystal. A sharp transition is clearly seen in *C/T* near 29 K. For further analysis of the superconducting specific heat we extrapolate the normal state background as indicated by the dashed line. Although this procedure is not suitable for determining the electronic specific heat at temperatures far below $T_c$, in the temperature range near $T_c$ the shape of the specific heat anomaly and its field dependence can be extracted reliably. For the heavily over doped samples (x = 0.45, 0.50, and 0.55) a field of 7.9 T || c suppressed superconductivity sufficiently so that these in-field data served as the normal state background.

Figure 2 shows the evolution of the transition in the electronic specific heat of the x = 0.3 and x = 0.5 crystal in various fields applied parallel to the *c*-axis and parallel to the *ab*-planes, respectively. In zero field the specific heat anomaly displays a sharp superconducting transition with a width of ~1 K which in increasing fields systematically shifts to lower temperatures and broadens weakly. The small broadening in applied fields is consistent with weak superconducting fluctuation effects as quantified by the low value of the Ginzburg number $G_i = \left(k_B \mu_0 \Gamma T_c / 4\pi \xi_{ab}^3 B_c^2\right)^2 / 2 = 323.6 \left(\Gamma \kappa \lambda_{ab} T_c\right)^2 \sim 5*10^{-5}$ (T in Kelvin, penetration depth $\lambda_{ab}$ in meters). The in-field behavior of the specific heat anomaly of $BaFe_2(As_{1-x}P_x)_2$ is analogous to that of optimally doped $Ba_{1-x}K_xFe_2As_2$ for which $G_i \sim 10^{-3}$ [16] has been estimated, but is in contrast to the pronounced broadening and suppression of the transitions seen in the more anisotropic $SmFeAsO_{0.85}F_{0.15}$ and $NdFeAsO_{0.82}F_{0.18}$ compounds with $G_i \sim 10^{-2}$ [17, 18].

A comparison of the H || c and H || ab-data reveals a superconducting anisotropy of $\Gamma \sim 2.6$, as directly inferred by superimposing the 2T || ab and 5T || c data (inset in Fig. 2a,b). We use an entropy conserving construction as indicated by the solid lines in Fig. 2a and 2c for determining the transition temperature $T_{c2}(H)$ and the height of the specific heat anomaly $\Delta C/T_c$. For the near optimally doped crystal (x = 0.30) we find $\Delta C/T_c$ = 54 mJ/mol K$^2$ whereas for a crystal with $T_c \sim$

28 K, $\Delta C/T_c$ = 45 mJ/mol K$^2$, which is in good agreement with a value of 38.5 mJ/mol K$^2$ obtained on an assembly of crystals with similar $T_c$ [6]. With increasing phosphorous doping $\Delta C/T_c$ drops significantly to 20 mJ/mol K$^2$ for the x = 0.5-crystal (see below) (Fig. 2c). In the temperature and field range covered here the upper critical field lines, $H_{c2}(T)$, are linear. With the help of the thermodynamic and Ginzburg-Landau relations $\mu_0 \Delta C/T_c = (\mu_0 dH_c/dT)^2_{T_c} = (\mu_0 dH_{c2}/dT)^2_{T_c} / \beta_A (2\kappa^2 - 1)$, $\kappa = \lambda_{ab}/\xi_{ab}$, and $\mu_0 H_{c2} = \phi_0/2\pi\xi^2$ we evaluate the doping dependence of materials parameters such as $\kappa$, coherence length $\xi_{ab}$, and Ginzburg-Landau penetration depth $\lambda_{GL}^{ab}$ as listed in Table 1.

Fig. 3 shows the dependence of the anisotropy parameter $\Gamma$ and of $\kappa_c$ upon $T_c$. Within the experimental uncertainty the anisotropy is independent of $T_c$, i. e., doping level, at a value of ~ 2.6. This independence is unexpected since band structure calculations [19] as well as dHvA measurements [7, 8, 20] indicate that the outer hole band develops increased c-axis dispersion implying a reduced anisotropy upon doping with P. Our result can be explained if the electron and inner hole bands - which do not change appreciably upon doping – dominate in forming the superconducting state. ARPES experiments on BaFe$_2$(As$_{0.7}$P$_{0.3}$)$_2$ [13] revealed almost isotropic superconducting gaps of 6 – 8 meV in the center plane of the Brillouin zone on all five Fermi surface sheets and strong $k_z$-dispersion of the gap on the α-hole Fermi surface suggesting a circular node near the top of the Brillouin zone. Such gap structure may account for the observed doping-independence of the anisotropy parameter, even though the detailed gap structure of BaFe$_2$(As$_{1-x}$P$_x$)$_2$ is still controversial [10,13]. In contrast to $\Gamma$, the c-axis GL-parameter decreases significantly upon doping; though the material remains clearly type-II for all P-concentrations (see discussion below).

Fig. 4 shows the doping dependence of $\Delta C/T_c$ and $-\mu_0 dH_{c2}^c/dT$. Both quantities are found to be approximately proportional to $T_c^2$ as indicated by the dotted lines. In a simple analysis, $-\mu_0 dH_{c2}^c/dT\big|_{T_c} \sim \phi_0/2\pi\xi^2(0)T_c$ which yields for the dirty limit $-\mu_0 dH_{c2}^c/dT\big|_{T_c} \sim \phi_0/2\pi v_F l$ and $-\mu_0 dH_{c2}^c/dT\big|_{T_c} \sim \phi_0 T_c/2\pi v_F^2$ in the clean limit. The observation of dHvA-oscillations indicates that at least the over-doped samples are close to the clean limit. Neither of these relations agrees with the observed $T_c^2$-variation; however, in addition to the explicit $T_c$-dependence there may arise unknown implicit $T_c$-dependences due to the doping dependence of Fermi velocity $v_F$ and mean free path $l$. In the limit of strong pair-breaking scattering and gapless superconductivity, $-\mu_0 dH_{c2}^c/dT \sim T_c$ has been predicted [21].

The jump height of the specific heat follows a $T_c^2$-relation reasonably well. In fact, the absolute size of $\Delta C/T_c$ agrees well with the trend observed for a large variety of Fe-pnictide / chalcogenide superconductors as shown in Fig. 5. Even though there may be appreciable deviations in this log-log plot, particularly at low values of $T_c$ and $\Delta C/T_c$, and additional uncertainties may arise due to varying sample quality and approximations made to extract the jump height [22], a trend extending over a large number of 122-, 111-, 11-based samples is clearly seen. It is remarkable that the $\Delta C/T_c$ vs. $T_c^2$ correlation is observed for electron, hole and isovalent doped (such as BaFe$_2$(As$_{1-x}$P$_x$)$_2$ presented here) materials. In addition, the BNC scaling occurs for under-doped and over-doped compounds as well as in clean samples, i.e. BaFe$_2$(As$_{1-x}$P$_x$)$_2$, and in strongly scattering charge doped compounds. A notable exception from this scaling is Sm-1111 [17] for which the specific heat anomaly is roughly a factor of 10 too small as compared to the BNC-scaling.

Assuming that the $\Delta C/T_c$ vs. $T_c^2$ line indeed represents a general property of a large class of Fe-based superconductors, then our results on $BaFe_2(As_{1-x}P_x)_2$ put constraints on models that have been advanced to account for the scaling. For instance, recognizing that within the conventional weak-coupling BCS-framework the $\Delta C/T_c$ vs. $T_c^2$ scaling implies a very peculiar balance of materials parameters, Zaanen [23] suggested non-Fermi liquid behavior in order to account for the wide occurrence of this scaling. However, dHvA-experiments [7, 8] on over-doped $BaFe_2(As_{1-x}P_x)_2$ yielded a Fermi surface topology in reasonable agreement with bandstructure calculations implying that at least this material can be described as Fermi liquid. Furthermore, Kogan [21] has pointed out that the $T_c^2$-variation can arise when strong pair-breaking scattering induces a gap-less superconducting state. Pair-breaking scattering could arise naturally in superconductors with sign-changing order-parameter such as $s^{\pm}$ gap symmetry due to non-magnetic inter-band scattering [24]. In the case of $BaFe_2(As_{1-x}P_x)_2$, however, the over-doped samples have reduced $T_c$ and $\Delta C/T_c$, but at the same time enhanced electron mean-free paths, that is, reduced scattering [7, 8, 19, 25], which is at odds with the theoretical model. Vavilov et al. [26] developed a model for the specific heat of under-doped compounds based on the coexistence of SDW and superconducting order. This model, however, does not account for the data in the over-doped regime. Recent specific heat measurements [27, 28] on $Ba(Fe_{1-x}Co_x)_2As_2$ revealed sizable residual terms $\gamma_r$ in the low temperature specific heat, that is, a contribution to the electronic specific heat arising from un-paired electrons, which shows a clear anti-correlation with $T_c$ and $\Delta C$: $\gamma_r$ is minimal at optimum doping and increases significantly on under and over doping. The entropy loss due to $\gamma_r$ could account for the rapid decrease of $\Delta C/T_c$ away from optimum doping. The origins of the $\gamma_r$-term have not been clarified yet. Since non-magnetic inter-band scattering in a $s^{\pm}$-superconductor is pair-breaking [24], scattering off charged dopant sites could induce quasi-particle states in the gap contributing to a residual $\gamma_r$ term. However, it is not obvious why this process would lead to a dependence of $\gamma_r$ and $\Delta C/T_c$ that is non-monotonous in Co-concentration. Alternatively, a "Swiss cheese" model has been invoked [27] in which a region of size of the coherence length around a defect site is driven normal thereby contributing to the residual electronic specific heat. This model has been applied to high-$T_c$ [29] and Heavy Fermion [30] materials. Since the coherence length is typically the smallest for the highest $T_c$, that is, at optimum doping, a non-monotonous variation of $\gamma_r$ with dopant concentration could arise.

For the optimally doped compound we obtain a Ginzburg-Landau penetration depth (linear extrapolation of $1/\lambda^2$ to zero-temperature) of $\lambda_{GL}^{ab}(0) \sim 101$ nm. Assuming the empirical two-fluid temperature dependence would yield a zero-temperature penetration depth of $\lambda_{ab}(0) = 2 \lambda_{GL}^{ab}(0) \sim 200$ nm, which is in good agreement with measurements based on micro-wave surface resistance [12]. With increasing P-doping, that is, decreasing $T_c$, the penetration depth increases implying a decrease of the lower critical field $H_{c1}$. This appears as a natural result; however, it has been noted [31] that with doping the effective mass $m^*$ on the electron Fermi surface sheets decreases [7] and at the same time the carrier concentration $n$ (of one polarity) increases and that therefore the superfluid density $n/m^* \sim 1/\lambda^2$ would increase, corresponding to a decrease of $\lambda$. This apparent contradiction may arise from the fact that our measurements yield values of the slopes of $H_{c2}$ and $1/\lambda^2$ near $T_c$ where GL-theory applies and that due to multi-band effects the actual temperature dependence of $\lambda$ deviates from the two-fluid form as evidenced by a pronounced inflection point in $1/\lambda^2(T)$ [6, 32]. A quantitative evaluation of the doping dependence of the low-temperature penetration depth will require the determination of the Fermi surface averages of the gap function and of the Fermi velocity on all sheets of realistic Fermi surfaces. We note though that due to the strong decrease of $\kappa_c$ with decreasing $T_c$, the relative increase of $\lambda_{GL}^{ab}$ is significantly smaller than that of the coherence length.

In conclusion, we present a systematic study of the specific heat transitions on a series of $BaFe_2(As_{1-x}P_x)_2$ single crystals with phosphorous doping ranging from near optimum doped x = 0.3 to strongly over-doped x = 0.55. $BaFe_2(As_{1-x}P_x)_2$ is the first member of the Fe-based layered superconductors that follows the BNC-scaling of the specific heat transition remarkably and - at the same time – can be made with very high purity. The high purity of this material imposes new restraints on theories aimed at explaining the scaling since models based on specific electron scattering mechanisms appear to be inconsistent with our results. Furthermore, we find that the Ginzburg-Landau parameter $\kappa_c$ decreases significantly with doping whereas the superconducting anisotropy is $\Gamma \sim 2.6$, independent of doping, indicative of the dominating role of the electron Fermi surface sheets in forming the superconducting state.

Crystal synthesis was supported by the Center for Emergent Superconductivity, an Energy Frontier Research Center funded by the US Department of Energy, Office of Science, Office of Basic Energy Sciences (CC, LF, WKK), materials characterization was supported by the core research program of the Department of Energy, Office of Science, Office of Basic Energy Sciences (UW, HC, GWC), under Contract No. DE-AC02-06CH11357. We acknowledge helpful discussions with V. Kogan, T. Shibauchi and M. Graf.

Figure and table captions

Table 1. Compilation of the superconducting parameters of samples with various $T_c$. The values for $\xi_{ab}(0)$ and $\lambda_{GL}^{ab}(0)$ are obtained from the measured upper critical field slopes using a linear extrapolation to zero-temperature according to $1/\xi_{ab}^2(0) = -2\pi T_c/\phi_0 \, dH_{c2}/dT$ and $1/[\lambda_{GL}^{ab}(0)]^2 = 1/[\kappa^2 \, \xi_{ab}^2(0)]$.

Fig.1. (Color online) lower inset: Temperature dependence of the magnetization of the crystals measured in 1 G || c after cooling in zero field. Top inset: Picture of a cleaved single crystal displaying a shiny and flat surface.

Fig. 2. (Color online) Temperature dependence of the eletronic specific heat of $BaFe_2(As_{0.7}P_{0.3})_2$ (a, b) and $BaFe_2(As_{0.5}P_{0.5})_2$ (c, d) near the superconducting transition in various fields applied parallel to the c axis and parallel to the *ab* plane. The solid lines in (a) illustrate the entropy conserving construction used to determine the transition temperature.

Fig. 3. (Color online) Dependence of the superconducting anisotropy parameter $\Gamma$ and the c-axis Ginzburg-Landau parameter $\kappa_c$ on $T_c$.

Fig. 4. (Color online) Dependence of the upper critical field slope and of the specific heat anomaly on $T_c^2$.

Fig. 5. (Color online) Compilation of $\Delta C/T_c$ vs. $T_c$ values of various Fe-based superconductors plotted on log-log scales. The line indicates a quadratic dependence. See also [5, 6, 22].

| $T_c(K)$ | x | $\mu_0 dH_{c2}^c/dT$ (T/K) | $\mu_0 dH_{c2}^{ab}/dT$ (T/K) | $\Gamma$ | $\Delta C/T_c$ (mJ/molK$^2$) | $\mu_0 H_{c2}^c$ (T) | $\kappa$ | $\xi_{ab}$(nm) | $\lambda_{GL}^{ab}$(nm) |
|---|---|---|---|---|---|---|---|---|---|
| 29.2 | 0.30 | -2.44 | -6.08 | 2.49 | 54.0 | 71.7 | 47.3 | 2.14 | 101 |
| 28.1 | 0.33 | -2.23 | -5.73 | 2.57 | 45.0 | 62.7 | 47.3 | 2.29 | 108 |
| 21.5 | 0.45 | -1.28 | -3.51 | 2.74 | 22.0 | 27.5 | 38.4 | 3.46 | 132 |
| 18.0 | 0.50 | -0.90 | -2.33 | 2.58 | 20.0 | 16.4 | 28.6 | 4.50 | 129 |
| 9.2 | 0.55 | -0.38 | -1.00 | 2.70 | 8.0 | 3.5 | 19.0 | 9.71 | 184 |

Table 1

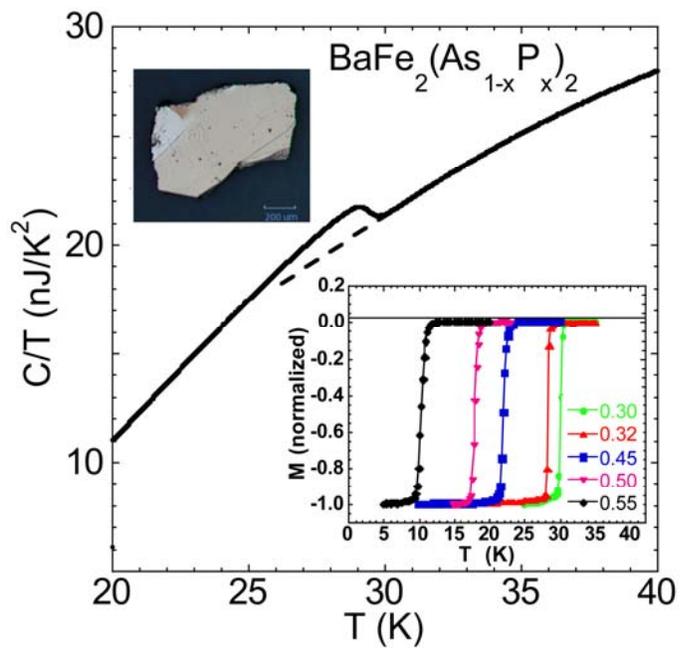

Fig. 1

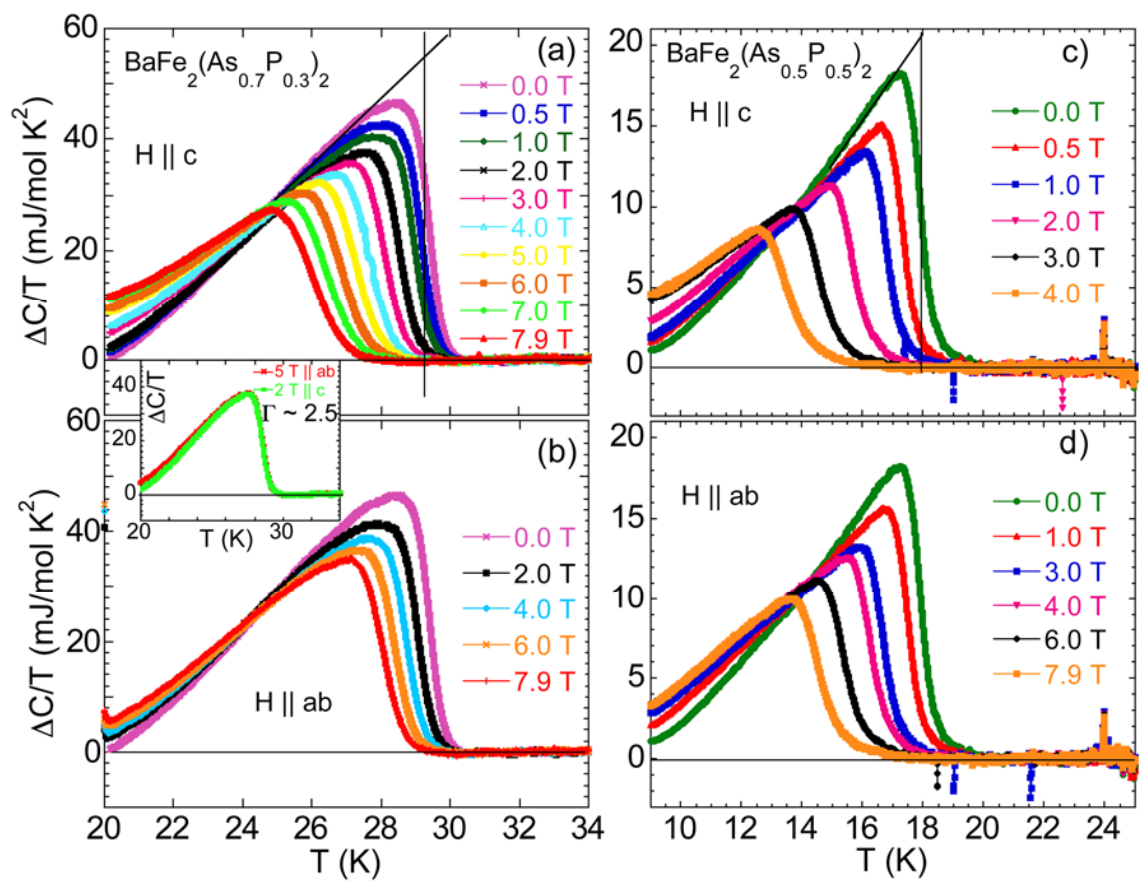

Fig. 2

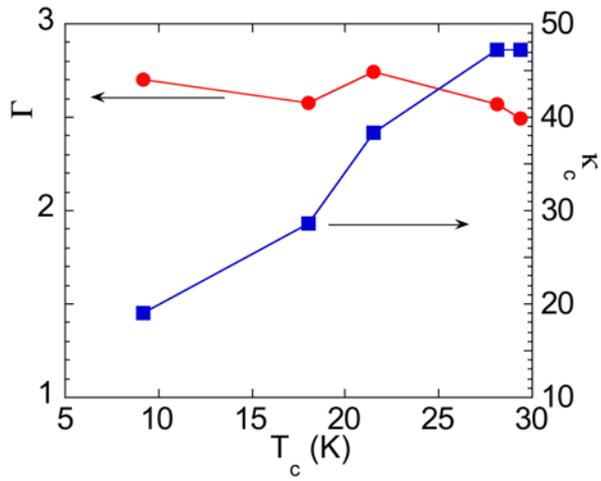

Fig. 3

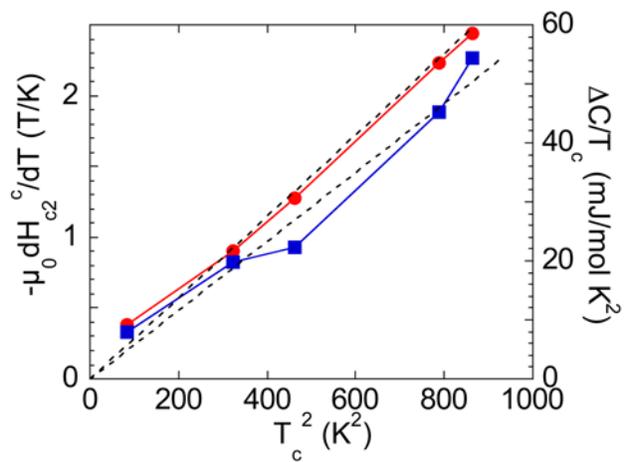

Fig. 4

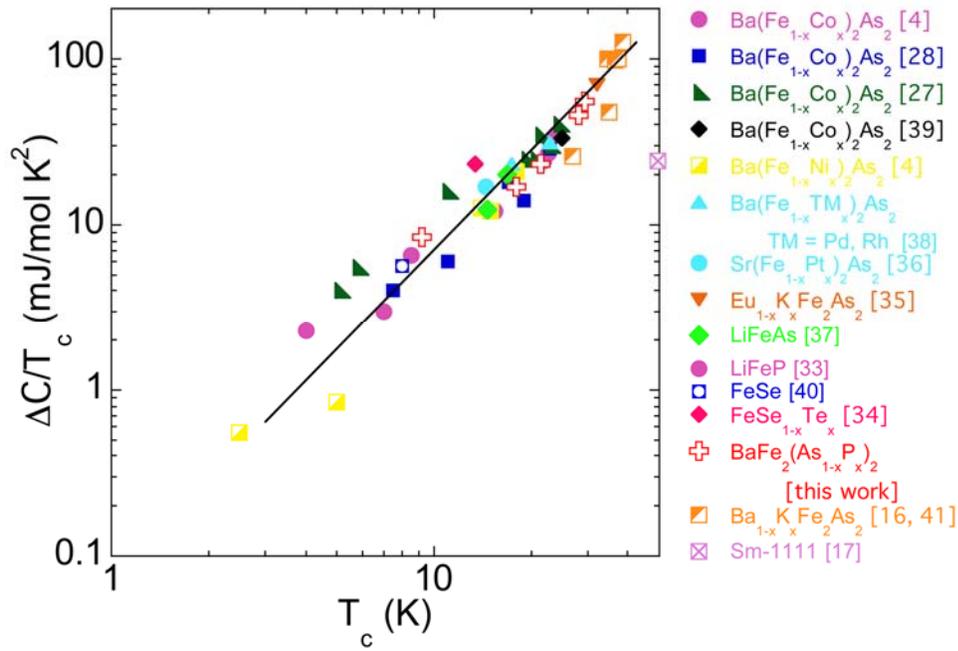

Fig. 5